\documentclass[manuscript]{aastex}

\shorttitle{GrayStarServer}
\shortauthors{Short}

\begin{document}


\title{GrayStarServer: Server-side spectrum synthesis with a browser-based client-side user interface}


\author{C. Ian Short}
\affil{Department of Astronomy \& Physics and Institute for Computational Astrophysics, Saint Mary's University,
    Halifax, NS, Canada, B3H 3C3}
\email{ian.short@smu.ca}




\begin{abstract}

We present GrayStarServer (GSS), a stellar atmospheric modeling and spectrum synthesis code
of pedagogical accuracy that is accessible in any web browser on commonplace computational devices
 and that runs on a time-scale of a few seconds.  The addition of spectrum
synthesis annotated with line identifications extends the functionality and pedagogical 
applicability of GSS beyond that of 
its predecessor, GrayStar3 (GS3).  The spectrum synthesis is based on a line list
acquired from the NIST atomic spectra database, and the GSS post-processing and user interface (UI) 
client allows
the user to inspect the plain text ASCII version of the line list, as well as to 
apply macroscopic broadening.  
Unlike GS3, GSS carries out the physical modeling on the server side
in Java, and communicates with the JavaScript and HTML client  
via an asynchronous
HTTP request.  We also describe other improvements beyond GS3 such as a more
physical treatment of background opacity and atmospheric physics, the comparison of key 
results with those of the Phoenix code \citep{phoenix}, and the use of the
HTML $<$canvas$>$ element for higher quality plotting and rendering of results.  We also present
LineListServer, a Java code for converting custom ASCII line lists in NIST format
to the byte data type file format required by GSS so that users can prepare their
own custom line lists.  We propose a standard for marking up and packaging model atmosphere
and spectrum synthesis output for data transmission and storage that will facilitate
a web-based approach to stellar atmospheric modeling and spectrum synthesis.
We describe some pedagogical demonstrations and exercises enabled by easily accessible,
 on-demand, responsive spectrum synthesis.  GSS may serve as a research support tool
by providing quick spectroscopic reconnaissance.  GSS may be found at
www.ap.smu.ca/$\sim$ishort/OpenStars/GrayStarServer/grayStarServer.html, and
source tarballs for local installations of both GrayStarServer and LineListServer
may be found at www.ap.smu.ca/$\sim$ishort/OpenStars/.    
   
\end{abstract}



\section{Introduction}

Understanding stellar atmospheres and their corresponding spectra has long been a corner stone in
both astronomy research and education.  Current areas of active
research that rely on synthetic stellar spectra include population synthesis of spatially
unresolved extragalactic populations, extracting the spectral signatures of planetary 
atmospheres from the light of their host stars, and automated stellar parameter estimation pipelines
for large spectroscopic surveys (see the Introduction of \citet{kirby}).
Research-grade stellar atmospheric modeling and spectrum synthesis is one of the most
technically complex and forbidding, and computationally demanding, procedures in 
astrophysical research, and does not lend itself to casual use in undergraduate teaching and 
learning on commonplace computational devices.
At the same time, the stellar atmospheres and spectra unit of the undergraduate curriculum
is the crucial first point of contact for many students with generally applicable important
ideas such as hydrostatic and thermal equilibrium, ionization and excitation equilibrium, 
radiative transfer and extinction (opacity), and spectral line formation.  Among the standard learning 
goals of the curriculum are
the understanding of the physical basis for the MK spectral class-$T_{\rm eff}$-relation,
the MK luminosity class-$\log g$ relation, and the spectroscopic determination of
abundance.  A natural way to untangle the complex entanglement of physical principles 
that determine these phenomena and relationships is parameter perturbation, and 
that suggests that didacticized computational modeling should be particularly useful for both
classroom demonstration and lab-style homework projects. 

\paragraph{}

In \citet{graystar3} and other papers in that series we present an approximate stellar atmosphere and
spectral line modeling code, GrayStar3 (GS3) in its most recent version, written in JavaScript (JS) that 
runs in any web browser, and that has a didactic
pedagogical user interface (UI) implemented in HTML.  Details about the computational methods employed can be
found in \citet{graystar3} and \citet{methods}, and suggested classroom demonstrations and lab-style 
projects can be found in \citet{graystar3} and \citet{pedagogy}.  The strengths of GS3 include the rendering
and display directly in the web browser of direct observables that can be qualitatively appreciated by a wide audience, 
such as the spatially 
resolved limb-darkened and -reddened white-light and tunable monochromatic 
disk images, and the visible flux spectrum with absorption from 14 important spectral lines, as well 
as numerous optionally displayable technical graphs of important modeled quantities.  These are created
by scripting HTML elements according to the results of internally self-consistent, albeit approximate, 
{\it in situ} physical modeling
in JS.  This provides a teaching and learning tool that is adaptable for a wide range of education and 
public outreach (EPO) levels, from the high school and general public level through to the introductory
graduate level.  GS3 allows stellar astronomy and astrophysics instructors at all levels to make use of 
physics education research (PER)-based demonstration-centered methods 
({\it eg.} see \citet{knight} and \citet{mazur}) in the classroom and to give students
hands-on lab-style homework projects. 

\paragraph{}

 Computational modeling in JS has limitations that restrict the modeling GS3 can perform.  JS code runs on
the machine on which the browser session is taking place (the ''client''), and for security reasons it does not
provide for file I/O.  Therefore, there is no means to separate large amounts of physical input data,
such as that comprising an atomic line list, from the source code.  Additionally, web browser programmers 
did not anticipate that JS would be used for intensive applications, and browsers become unresponsive
and browser sessions unrecoverable if a JS script exceeds a brief execution time (typically 20 to 30
seconds) (although some common browsers available for the Windows 8/10 OS seem less prone to this 
than others).  As a result, GS3 was limited to including extinction from only 14 of the most 
pedagogically important MK classification and Fraunhofer lines in its computation of the emergent
stellar spectrum.  This is sufficient to address some topics in MK spectral classification, but does
not allow for genuine {\it spectrum synthesis}, in which a spectral region is computed with incorporation 
of most or all the line extinction that is significant.  A particularly unfortunate consequence is 
that the Johnson {\it U-B} and {\it B-V} photometric color indices computed by GS3 do not reflect the influence
of line blanketing on the near UV and blue part of the spectral energy distribution (SED), and are 
significantly discrepant with observed values for a star of given {\it V-R} index value.     
Another limitation of GS3 is that, as an exploratory proof of concept application,
the modeling incorporated crude estimates of the electron pressure ($P_{\rm e}(\tau)$), 
mean molecular weight ($\mu(\tau)$), and wavelength-averaged mean extinction ($\kappa(\tau)$), structures
from simple re-scaling of research-grade models of the Sun or Vega, 
and made use of a wavelength-independent gray extinction in its calculation of the spectral line
profile.  As a result, the geometric depth scale was often compressed or expanded by a factor of 10 or more 
compared to research-grade modeling results for models that differed significantly from either the Sun or Vega, 
and the strength of many spectral lines, 
as expressed as equivalent width, $W_\lambda$,
was often discrepant with observations and research-grade modeling results by a factor of a few.            

\paragraph{}

  To overcome these limitations, we have developed GrayStarServer (GSS), which performs the physical
modeling on the server, without the restrictions of client-side processing in JS, while still
taking input and displaying results in a GS3-like client-side web browser-based UI.  This allows GSS to
be a pedagogical spectrum synthesis code as well as an atmospheric modeling code.
Moreover, the crude modeling of GS3 has been replaced by a much more proper, physically
correct treatment that gives results that are closer to that of research-grade modeling 
than that of GS3 generally throughout the HR diagram.  The modeling methods are not new, and we 
emphasize that the novel contribution is that the modeling and visualization are taking
place is a platform-independent web-deployment framework.   
GSS may be found at www.ap.smu.ca/$\sim$ishort/OpenStars/GrayStarServer/grayStarServer.html.  Users
who wish to have their own local installation may find source tarballs for GSS and for the
accompanying line list utility, LineListServer (see below), at www.ap.smu.ca/$\sim$ishort/OpenStars/.
In Section \ref{Server} we describe the server-side code, including improvements with respect to GS3,
compare key results with those of the well-known research-grade code Phoenix,  
and describe the interaction with the client-side UI; in Section \ref{Client} 
we describe the client-side post-processing package and UI with
emphasis on the contrast with the GS3 UI; in Section \ref{ClientServer} we discuss special
issues related to the client-server interaction and propose a standard for packaging and
marking up stellar atmospheric and spectrum synthesis data for transmission with HTTP;  
Sections \ref{EPO} and \ref{Research} discuss the applications
to EPO and to research, respectively.  In Section \ref{Conclude} we present conclusions.

\section{Server-side modeling in Java \label{Server}}

\paragraph{}

  All computational modeling is carried out on the machine that hosts the GSS web site (the ''server''),
including both the atmospheric structure and the spectrum synthesis calculations, and the code has been 
written in the Java programing language.  

\subsection{Improved atmospheric structure modeling \label{newphysics}}

In GS3, and in its Java development version, GrayFox, some of the key distributions that determine 
the model structure and the strength or shape of 
spectral lines, such as the $P_{\rm e}(\tau)$, $\mu(\tau)$, and the wavelength-independent gray
$\kappa(\tau)$ structures, were crudely estimated from re-scaling from the results of a research-grade model
of either the Sun or Vega (A0 V), and the gray $\kappa(\tau)$ value was being used for the
background continuous extinction, $\kappa^{\rm c}(\tau)$, in the line profile calculation
\citep{methods}, \citep{graystar3}.
GSS improves greatly on the situation by following the procedure outlined in Chapters 8
and 9 of \citet{dfg3} to compute $\lambda$-dependent physics-based cross-sections, $\sigma_\lambda$,
and the resulting extinction contribution, $\kappa_\lambda(\tau)$,
for all the photon processes of H, He, or free electrons ($e^-$) at temperatures 
found in stellar atmospheres anywhere in the HR diagram, and to then compute the
atmospheric structure from the coupled structure equations, with the
exception of the thermal equilibrium equation.

\paragraph{}

\begin{figure}
\plotone{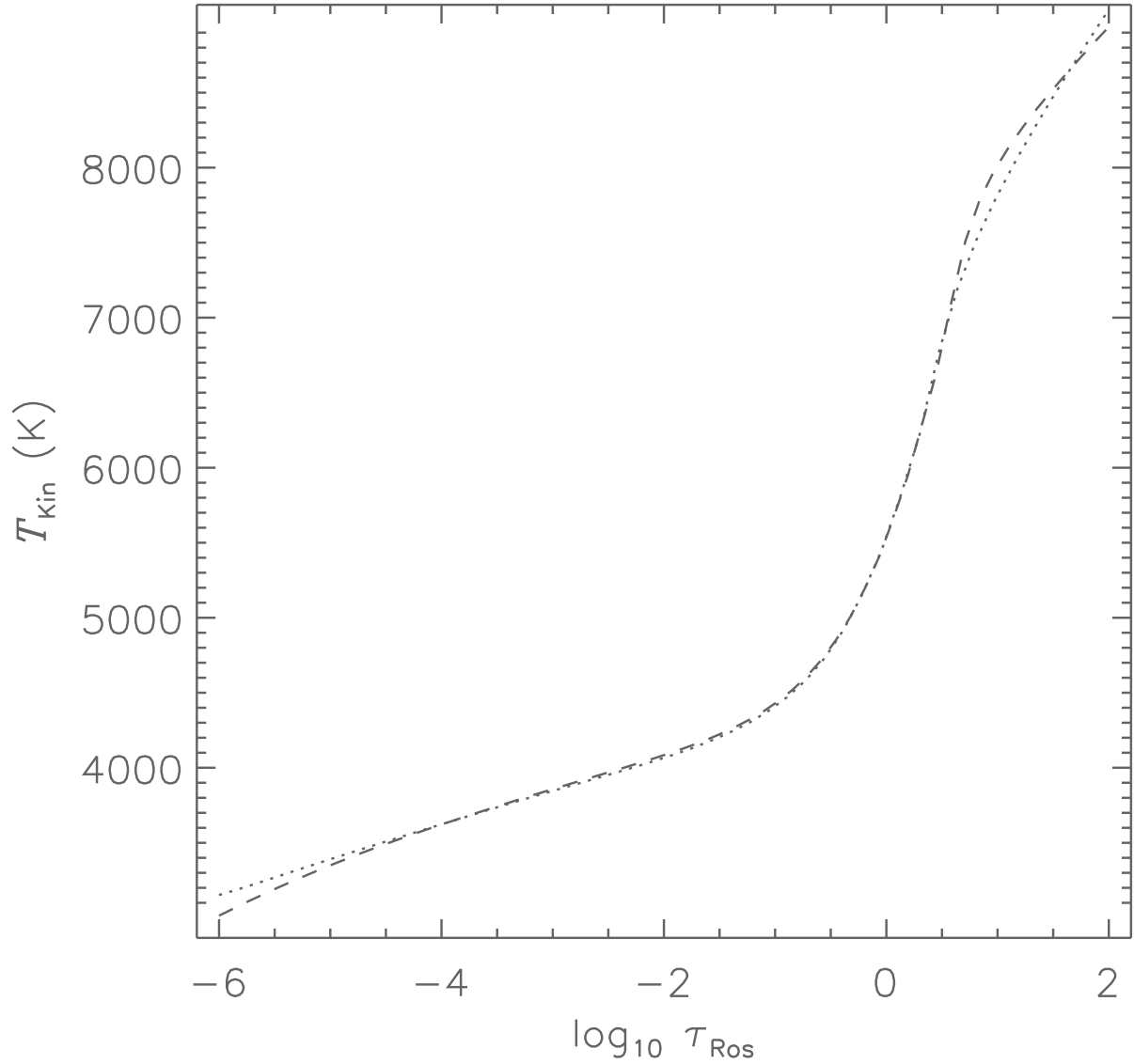}
\caption{ The kinetic temperature, $T_{\rm kin}(\tau)$, structure for a model of 
 $T_{\rm eff}=5000$ K, $[{{\rm Fe}\over {\rm H}}]=0.0$, and $\log g = 2.5$ as computed exactly 
by Phoenix (dotted line), and as re-scaled from a reference model of 
$T_{\rm eff}=5000$ K, $[{{\rm Fe}\over {\rm H}}]=0.0$, and $\log g = 4.5$ by GSS (dashed line).
  \label{gssTemp5000}
}
\end{figure}

  The procedure is outlined below.  Wherever possible, expressions are evaluated logarithmically,
and quantities are renormalized before direct subtraction.  Numerical integration is generally
performed with the extended trapezoidal rule, accurate to second order ($\mathcal{O}((\Delta x)^2)$), 
for simplicity.  Because a key aspect of the GrayStar project
is that the code is open source and in the public domain, we identify the corresponding routines
using the Java Class.method() notation:

\begin{enumerate}

\item{We establish an {\it ad hoc} optical depth grid, $\{\tau_{\rm i}\}$, with 48 points 
distributed uniformly in $\log_{\rm 10}\tau$ from -6 to 2 (six $\tau$-points
per decade).  All structures that are re-scaled from reference models (see below) are
interpolated onto this grid.  Method TauScale.tauScale().} 

\item{The kinetic temperature structure $T_{\rm kin}(\tau)$ is computed by re-scaling with $T_{\rm eff}$
the $T_{\rm kin}(\tau)$ structure of a research-grade {\it reference} model computed with Phoenix \citep{phoenix} of
either ($T_{\rm eff}$/$\log g$/$[{{\rm Fe}\over {\rm H}}]$/$\xi_{\rm T}$) equal to (5000 K/4.5/0.0/1.0 km s$^{\rm -1}$) - 
approximately a K1 V star,
or (10000 K/4.0/0.0/2.0 km s$^{\rm -1}$) - approximately a B9 V star, depending on whether the target model is a late-type
($T_{\rm eff} < 7300$ K) or an early-type ($T_{\rm eff} > 7300$ K) star.  At each value of
$\tau$, 
$T_{\rm kin}^{\rm Target}(\tau) = (T_{\rm eff}^{\rm Target}/T_{\rm eff}^{\rm Reference}) \times T_{\rm kin}^{\rm Reference}(\tau)$. 
GSS in its current public 
mode does not address the problem of thermal radiative-convective equilibrium and makes no further 
adjustment to $T_{\rm kin}(\tau)$.  As discussed in \citet{dfg3}, re-scaling with $T_{\rm eff}$
is surprisingly accurate, as can be seen to ''first order'' by considering the formula for
$T_{\rm kin}(\tau)$ from the gray approximation.  We perform this re-scaling separately for 
early- and late-type stars because their $T_{\rm kin}(\tau)$ structures differ because of the role
of convection and molecular opacity in the latter.  Fig. \ref{gssTemp5000} shows a comparison
of the $T_{\rm kin}(\tau)$ structures for a star of ($T_{\rm eff}$/$\log g$/$[{{\rm Fe}\over {\rm H}}]$/$\xi_{\rm T}$) 
equal to (5000 K/2.5/0.0/1.0 km s$^{\rm -1}$) - approximately a G6 III star - from applying this re-scaling, 
and as computed exactly with Phoenix.  Method phxRefTemp() in classes ScaleT10000 and ScaleT5000.}

\item{Starting approximations for the gas pressure, $P_{\rm gas}(\tau)$, and the partial electron
pressure, $P_{\rm e}(\tau)$, structures are produced by re-scaling the $P_{\rm gas}(\tau)$ and
$P_{\rm e}(\tau)$ structures of the relevant reference model with $\log g$,
$[{Fe\over H}]$, and He abundance, $A_{\rm He}$, according to the prescriptions given in
Chapter 9 of \citet{dfg3}.  The scaling with $\log g$ and with $[{{\rm Fe}\over {\rm H}}]$ for both 
$P_{\rm gas}$ and $P_{\rm e}$ is crudely temperature dependent in that the
exponent in the scaling law differs for early- and late-type stars.  The resulting
$P_{\rm gas}(\tau)$ and $P_{\rm e}(\tau)$ structures are to be refined by iteration.  
Methods phxRefPGas() and phxRefPe() in
classes ScaleT10000 and ScaleT5000.}

\item{The Hydrogen number density structure, $N_{\rm H}(\tau)$, is computed from the ideal
gas law equation of state (EOS) as 
$N_{\rm H}(\tau) = (P_{\rm gas}(\tau) - P_{\rm e}(\tau)) / kT_{\rm kin}(\tau)(\Sigma_{\rm Z}^{\rm Species}{A_{\rm Z}}$),
where $A_{\rm Z}$ is the fractional abundance of chemical element $Z$ with respect to H, 
$N_{\rm Z}/N_{\rm H}$, and we include all elements up to Zn (Z=30) and four relatively abundant neutron 
capture elements (Rb ($Z=37$), Sr ($Z=38$), Cs ($Z=55$), and Ba ($Z=56$)).  The number density structures 
of the remaining elements then follow 
from $N_{\rm Z}(\tau) = A_{\rm Z}N_{\rm H}(\tau)$, and the mass density structure follows
from $\rho(\tau) = \Sigma_{\rm Z}^{\rm Species}N_{\rm Z}\mu_{\rm Z}$, where $\mu_{\rm Z}$ is the
atomic mass of element $Z$.  We use the solar abundance distribution of \citet{grevesse10},
scaled by the user-input value of $[{{\rm Fe}\over {\rm H}}]$, for the $A_{\rm Z}$ values.
 Methods getNz() and massDensity2() in class State.}

\item{The ionization fractions, $f_{\rm k}(\tau)$, are computed for the first five ionization stages
($k = 1$ to 5, or fewer for H, He, Li, and Be) of all species, $Z$, using the last estimate of
$P_{\rm e}(\tau)$ on the right hand side of the Saha equation for $N_{\rm k+1}/N_{\rm k}$.  The $f_{\rm k}(\tau)$
structures are then used to refine the $N_{\rm e}(\tau)$ (and $P_{\rm e}(\tau)$) structures 
by accounting for $e^-$ particles liberated by ionization, as 
$N_{\rm e}(\tau) = \Sigma_{\rm Z}\Sigma_{k=2}^5{(k-1)f_{\rm k, Z}(\tau)N_{\rm Z}(\tau)}$.  The $f_{\rm k}(\tau)$
values and the $N_{\rm e}(\tau)$ structure are refined by three iterations of this step. 
(To ensure rapid responsiveness, we do not try to achieve convergence by any criterion.)
The refined $N_{\rm e}(\tau)$ structure is used to compute the mean molecular weight,
$\mu(\tau)$, structure.  Method LevelPopsServer.stagePops().}      

\item{The $T_{\rm kin}(\tau)$ and refined $P_{\rm e}(\tau)$ structures are used to 
compute the linear extinction coefficients, $\kappa_\lambda(\tau)$, in units of cm$^{-1}$
per neutral H atom, or per H particle, for eight extinction sources: \ion{H}{1} $b-f$
for the lowest 30 atomic energy levels,
\ion{H}{1} $f-f$, H$^-$ $b-f$, H$^-$ $f-f$, H$^{+}_{\rm 2}$ absorption, \ion{He}{1}
$b-f$ and $f-f$, He$^-$ $f-f$, and $e^-$ Thomson scattering.  Following the procedure in
Chapter 8 of \citet{dfg3}, the hydrogenic formulae for $\sigma_\lambda$ and the approximate formulae 
for the quantum mechanical Gaunt factors, $g_{\rm bf}$ and $g_{\rm ff}$, are used for 
the \ion{H}{1} $b-f$ and $f-f$ sources, the $\sigma_\lambda(T)$ values for the H$^-$ $b-f$
and $f-f$, H$^{+}_{\rm 2}$, and He$^-$ $f-f$ sources are based on polynomial fits to the 
relevant physical data,
and the total \ion{He}{1} $b-f$ and $f-f$ extinction is scaled approximately from the 
total \ion{H}{1} $b-f$ and $f-f$ extinction.  These are then put onto a common
linear extinction scale in units of cm$^{-1}$, added together at each $\tau$ point, 
and converted to continuum mass extinction
coefficients, $\kappa_\lambda^{\rm c}(\tau)$, by depth-wise division by the $\rho(\tau)$ structure. 
The Rosseland mean mass extinction coefficient, $\kappa_{\rm Ros}(\tau)$, is then
computed from the $\kappa_\lambda^{\rm c}(\tau)$ distribution.  We do not account for 
extinction from metals, or from molecules other than H$^{+}_{\rm 2}$, and we expect
our $\kappa_\lambda^{\rm c}(\tau)$ distribution to increasingly underestimate the
true value as $\lambda$ decreases below 400 nm.  Methods kappas2() and kapRos() in 
class Kappas. }

\item{The total pressure structure, $P(\log\tau)$, is then refined by integrating the 
formal solution of the hydrostatic equilibrium (HSE) equation on the logarithmic optical
depth scale, $\log\tau$, with an {\it ad hoc} initial condition of $\log P(\log\tau=-6)$ equal to
-4 $\log$ dynes cm$^{-2}$, which requires the $\kappa_{\rm Ros}(\tau)$ structure from the
previous step.  The bolometric radiation pressure, $P_{\rm rad}(\tau)$, is then computed 
under the assumption of a black-body intensity distribution ($I_\lambda(\tau)$ given by 
$B_\lambda(T_{\rm rad}=T_{\rm kin}(\tau))$), and 
the $P_{\rm gas}(\tau)$ structure is recovered by evaluating 
$e^f$, where $f$ is equal to the value of 
$\log P(\tau) + \log(1 - e^g)$, where $g$ is equal to the value of 
$\log P_{\rm rad}(\tau) - \log P(\tau)$.  To avoid unphysically low values of 
$P_{\rm gas}$ in the upper atmospheres of early-type stars, we artificially limit the value
of $P_{\rm rad}(\tau)/P(\tau)$ to 0.5 (in reality such stars are not in HSE in
the layers where this limit is applied).  Methods hydroFormalSoln() and radPress()
in class Hydrostat.}

\item{Steps 4 through 7 are iterated three times, each involving three sub-iterations
of step 5.  Again, to ensure responsiveness, we do not try to achieve a convergence
criterion.  Figs. \ref{gssPe5000} and \ref{gssPe10000} show the $P_{\rm gas}(\tau)$ and
$P_{\rm e}(\tau)$ structures resulting from this procedure as compared to
direct computation with Phoenix for models of 
($T_{\rm eff}$/$\log g$/$[{{\rm Fe}\over {\rm H}}]$/$\xi_{\rm T}$) equal to (5000 K/4.5/0.0/1.0 km s$^{\rm -1}$)
and (5000 K/2.5/0.0/1.0 km s$^{\rm -1}$), and of (10000 K/4.0/0.0/1.0 km s$^{\rm -1}$),
respectively.  Our approximate procedure yields 
results comparable with those of Phoenix research-grade modeling for early-type models, 
and for late-type of a range of $\log g$ values.  
}

\item{We compute the geometric {\it depth} scale, $z(\tau)$, defined as
increasing {\it in}ward, by re-arranging and integrating
the definition of the radial optical depth scale, ${\rm d}\tau(z) = \kappa_{\rm Ros}(z)\rho(z){\rm d}z$,
with an initial condition of $z(\tau_{\rm Ros}=-6)=0$ cm. 
 }

\end{enumerate}

The above procedure necessarily involves a subtle inconsistency in that the Phoenix reference
models are tabulated on the $\tau_{\rm 12000}$ scale (monochromatic continuous extinction
at 1200 nm), whereas we effectively interpret it as the Rosseland optical depth
scale $\tau_{\rm Ros}$.  However, the reference $P_{\rm gas}(\tau)$ and $P_{\rm e}(\tau)$
distributions are being taken as initial estimates that are subsequently refined, and 
any distortions in the $T_{\rm kin}(\tau)$ structure are minor compared to the
approximations of computing the $T_{\rm kin}(\tau)$ by re-scaling and neglecting metal
extinction, and, for late-type stars, neglecting molecular extinction and convection.

Because of the larger number of $\lambda$ points required by spectrum synthesis, the 
number of angles with respect to the local surface normal, $\theta$, at which the out-going
surface monochromatic specific intensity distribution, $I^{\rm +}_\lambda(\tau=0, \theta)$,
is sampled, is restricted to 11 (taken from the 
zero-positive domain of a 20-point Gauss-Legendre quadrature) rather
than the 17 currently used in GS3.  Using only 11 $\theta$ values for the outgoing quadrant 
under-samples $I^{\rm +}_\lambda(\theta)$ with respect to research-grade modeling, in which the 
value is more typically 16, but 
our choice crucially limits the number of ($\lambda$, $\theta$) pairs for which     
$I^{\rm +}_\lambda(\tau=0, \theta)$ must be computed, and is consistent with the overall
approximate nature of GSS as compared to research-grade modeling.  All other discretizations are
the same as in GS3 (except the $\lambda$ sampling of the spectrum, of course). 
In the description below we make use of the Java Class.method() notation to identify new
post-GS3 procedures in the GrayStarServer and LineListServer Java packages.  

\subsection{Spectrum synthesis}

\paragraph{}

  Beyond what GS3 does, GSS also computes a synthetic spectrum based on a significant atomic line list 
({\it ie.} spectrum synthesis).  The method employed is an extension of the method used in GS3 to
accommodate 14 lines, and represents an unusual approach for LTE spectrum synthesis to reduce
both computing time and the amount of data that must be transmitted from the server to the client.  

\paragraph{Line list}

  We create the line list from the ASCII plain text output from the NIST atomic spectra database
\citep{nist} (\url{http://physics.nist.gov/asd}).  The optional oscillator strength, $f_{\rm ij}$, 
and $\log gf$ fields are requested, as well as the standard excitation energy, $\chi_{\rm i}$ of the lower 
energy level, $i$, of the transition, and the line center wavelength in vacuum, $\lambda_0$.  The
line list processing procedure combines the $f_{\rm ij}$ and $\log gf$ values to extract the 
statistical weight, $g_{\rm i}$, of the level $i$.  As of this report, the line list
contains all permitted and forbidden lines in the database of $\log f_{\rm ij} > -5$ with
$300 < \lambda_0 < 900$ nm for all ionization stages up to and including stage IV for
all elements up to and including Zn ($Z=30$), and the first two ionization stages of four
relatively observable important neutron capture elements (Rb, Sr, Cs, and Ba), and amounts to just over $10^4$ 
lines.  The list includes all separately listed multiple components of \ion{H}{1} lines.   

\paragraph{}

 We have developed a Java package called LineListServer that reads in the NIST ASCII line list, 
performs preliminary processing, including the determination of $g_{\rm i}$ values, converts it to 
byte data type, and writes it to a machine readable byte file that GSS reads using the Java 
BufferedInputStream.read() method and the FileInputStream class.
The byte version of the line list is only 0.5 Mbytes
in size and is read very quickly as byte array data ($\sim 1$ second when using a buffer size
of 8 kbytes).

\paragraph{Energy level populations}

   For each species, $k$ (element and ionization stage), represented in the line list, GSS
initially computes the total species populations, $N_{\rm k}(\tau)$, once and for all with the Saha
equation, and stores it in an array held in memory.  
 The partition function of species $k$, $U_{\rm k}$, for the first two ionization stages
(I and II) is estimated by interpolation in 
temperature between the low- and high-temperature values given in \cite{allens}.  
The partition functions of the higher ionization stages 
is approximated with the ground state statistical weight for that
species, $g_{{\rm k, i=} 1}$, which is more accurate for lower than for higher $T$ values. 
  The lower energy level population of a line transition ($i\rightarrow j$), $n_{\rm i}$, is
computed on demand with the Boltzmann equation for each line in the line list.  This results 
in some instances of $n_{\rm i}$ being
computed more than once in the case of multiple lines having a common lower level, $i$.
However, the computation of $n_{\rm i}$ from $N_{\rm k}$ is done logarithmically and involves
only two additions, one multiplication, and one exponentiation when the denominator, 
$kT_{\rm kin}(\tau_{\rm Ros})$, in the Boltzmann factor has been precomputed.  This approach 
has the advantage of avoiding a scheme for matching each transition, $i\rightarrow j$, 
to a precomputed corresponding $n_{\rm i}$ value, and is suitable for 
our modest line list where the number of re-computations of $n_{\rm k, i}$ for the same
$k$ and $i$ is limited.  The LevelPops.levelPops()
method of GrayFox has been modified to provide the LevelPopsServer.stagePops() and
LevelPopsServer.levelPops() methods for computing the $N_{\rm k}$ and $n_{\rm i}$
values, respectively.

\paragraph{Wavelength sampling}

  GSS starts with the very coarse grid of 250 $\lambda$ points that is employed in GS3 to sample the overall SED 
uniformly in $\log\lambda$.  A pure continuum surface flux spectrum, $F^{\rm c}_\lambda$, that neglects line 
opacity is computed on
this sparse grid as in \citet{graystar3} for use in continuum rectification of the spectrum with line opacity.
For each spectral line that meets the line-to-continuum extinction 
($\kappa^{\rm l}/\kappa^{\rm c}$) criterion for
inclusion,
GSS inserts a grid of $\Delta\lambda$ points centered on the $\lambda_0$ value of the line
($\Delta\lambda = \lambda - \lambda_0$) that samples the
line profile uniformly in the Gaussian core and logarithmically in the Lorentzian wings.  
The GrayFox LineGrid.lineGridVoigt()
method is used to create the grid 
of $\Delta\lambda$ points.  
This results in a highly non-uniform, sparse $\lambda$ sampling of the spectrum synthesis
region in which $\lambda$ points are generally {\it only} present where these is a spectral line,
and continuum regions between lines are sparsely sampled.

\paragraph{}

GSS then computes the normalized line profile, $\phi_\lambda(\Delta\lambda)$, and the 
monochromatic {\it line} extinction coefficient distribution, $\kappa^{\rm l}_\lambda(\tau_{\rm Ros})$, 
for each line with the original GrayFox
LineProf.voigt() and LineKappa.lineKap() methods, respectively.  
Given our unusually sparse $\lambda$ sampling that largely covers spectral lines only, we treat
{\it all} lines with full approximate Voigt profiles ({\it ie.} including the Lorentzian wings),
including those lines that might be weak enough to normally warrant treatment of the Gaussian 
core only.  Line broadening is treated as in GS3 - all broadening mechanisms
are assumed to produce Lorentzian wings (including the non-Lorentzian $e^-$-impact linear Stark broadening
of \ion{H}{1} lines), characterized by a broadening parameter 
$\gamma_{\rm Lorentz}$, and the user may adjust a universal logarithmic 
Lorentzian broadening enhancement factor, $\gamma_{\rm extra}$, 
between values of 0 and 1.  Beyond the GS3
treatment reported in \citet{graystar3}, GSS accounts for radiation
(natural) broadening by adding the Einstein coefficient for spontaneous
de-excitation, $A_{\rm ij}$, taken from the NIST line list, such 
that 
$\gamma_{\rm Total} = \gamma_{\rm Lorentz}\gamma_{\rm extra} + A_{\rm ij}$.
This treatment underestimates the amount of radiation broadening by
neglecting the effect of downward transitions from level $j$ other than $j\rightarrow i$
on the life-time of the upper level $j$, but is consistent with the overall
approximate nature of GSS.

\paragraph{}

As GSS processes the line list,
the $\kappa^{\rm l}_{\lambda}(\tau_{\rm Ros})$ distribution for each successive line that is to be included
is added cumulatively to the total monochromatic extinction distribution, 
$\kappa_{\lambda}(\tau_{\rm Ros})$, so that line blending is approximately accounted for.  
The computation of the total monochromatic optical
depth scale, $\tau_\lambda$, and the 
$I^{\rm +}_\lambda(\tau=0, \theta)$ distribution, are then computed as is GS3
with the LineTau2.tauLambda() and FormalSoln.formalsoln() methods.

\paragraph{Line selection}

   For each line in the list that falls within the requested synthesis region, GSS performs an initial 
computation of the ratio of line extinction 
to continuous extinction at line center, $\kappa^{\rm l}(\lambda_0)/\kappa^{\rm c}(\lambda_0)$,
at three sample $\log\tau_{\rm Ros}$ values (-5.0, -3.0, -1.0),
and discards the line if the value is below the criterion for inclusion at {\it all three} depths.  This conforms
to standard practice established by research-grade codes.  The procedure is the
same as that for full line profiles, except that the new LineGrid.lineGridDelta() and LineProf.delta()
methods are used in place of the LineGrid.lineGridVoigt() and LineProf.voigt() methods to treat
the line with a single-point line profile.  To expedite the computation when the application first loads,
the default value of $\kappa^{\rm l}(\lambda_0)/\kappa^{\rm c}(\lambda_0)$ is -2.0 so that only 
significantly strong lines are accounted for, but the value may be reduced to a lower limit of -4.0.

\paragraph{}

\begin{figure}
\plotone{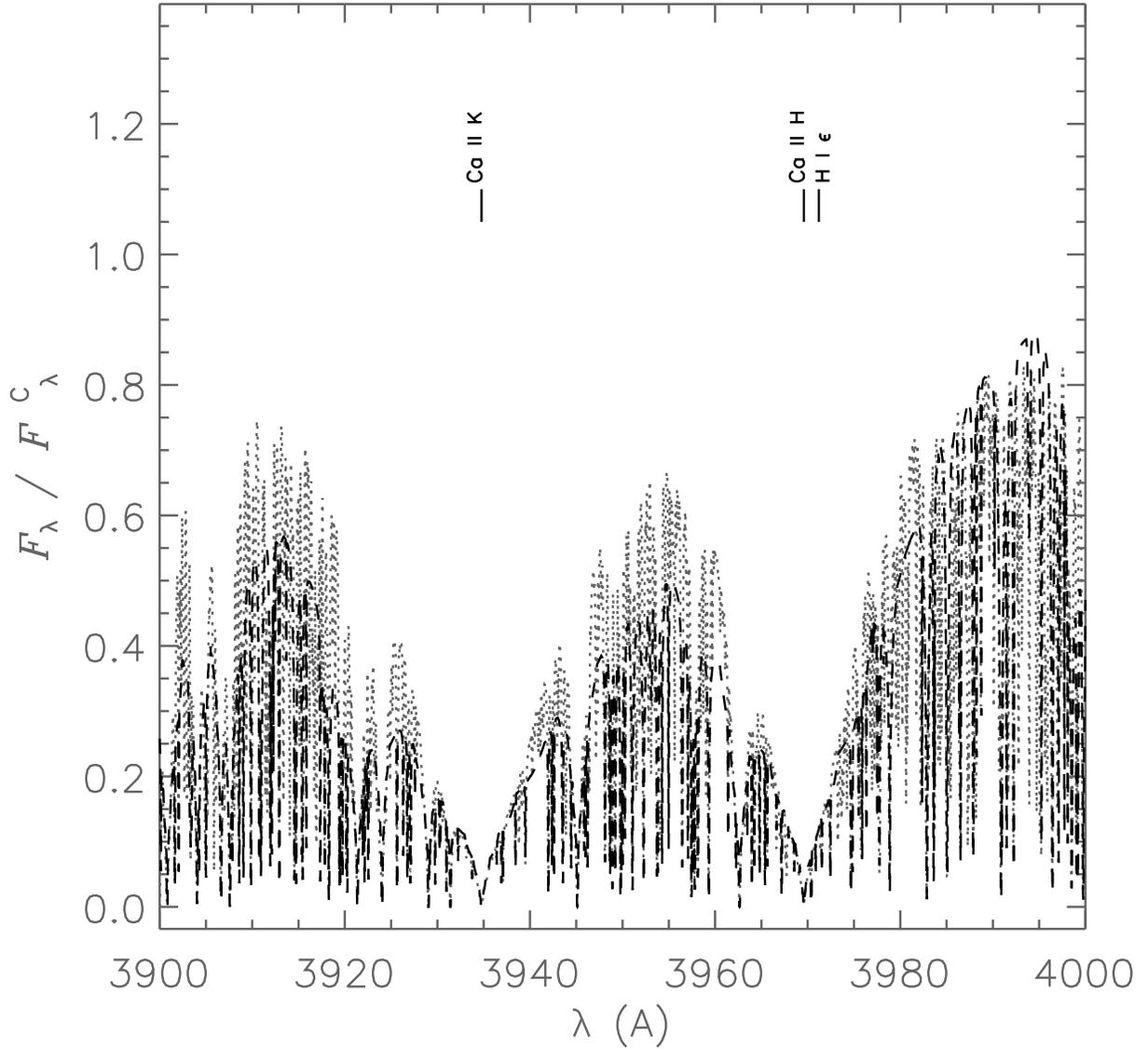}
\caption{Continuum rectified synthetic flux spectra, $F_\lambda(\lambda) / F_\lambda^{\rm c}(\lambda)$, in 
the important \ion{Ca}{2} $HK$  
region for a model of $T_{\rm eff}=5000$ K, $[{{\rm Fe}\over {\rm H}}]=0.0$, $\log g=4.5$,
and $\xi_{\rm T}=1.0$ km s$^{-1}$ as computed with GSS (dashed line) and 
Phoenix (dotted line).  
  \label{gssFlux500045}
}
\end{figure}

\begin{figure}
\plotone{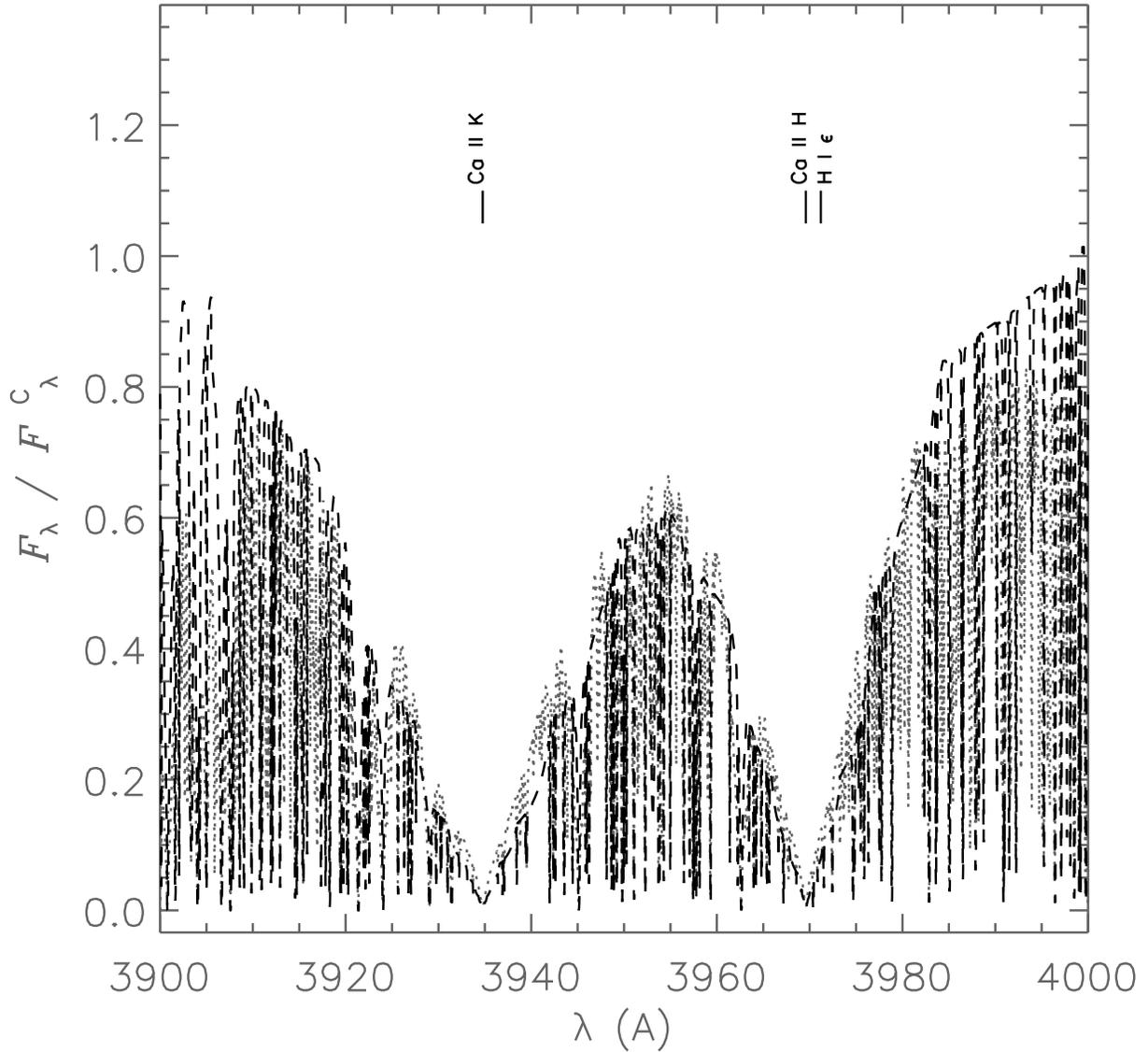}
\caption{ Same as Fig. \ref{gssFlux500045}, except  
for a model of $T_{\rm eff}=5000$ K, $[{{\rm Fe}\over {\rm H}}]=0.0$, $\log g=2.5$,
and $\xi_{\rm T}=1.0$ km s$^-1$.  Comparison with Fig. \ref{gssFlux500045} allows
a comparison of how GSS and Phoenix account for $g$-dependent 
pressure broadening of saturated spectral lines. 
  \label{gssFlux500025}
}
\end{figure}

\begin{figure}
\plotone{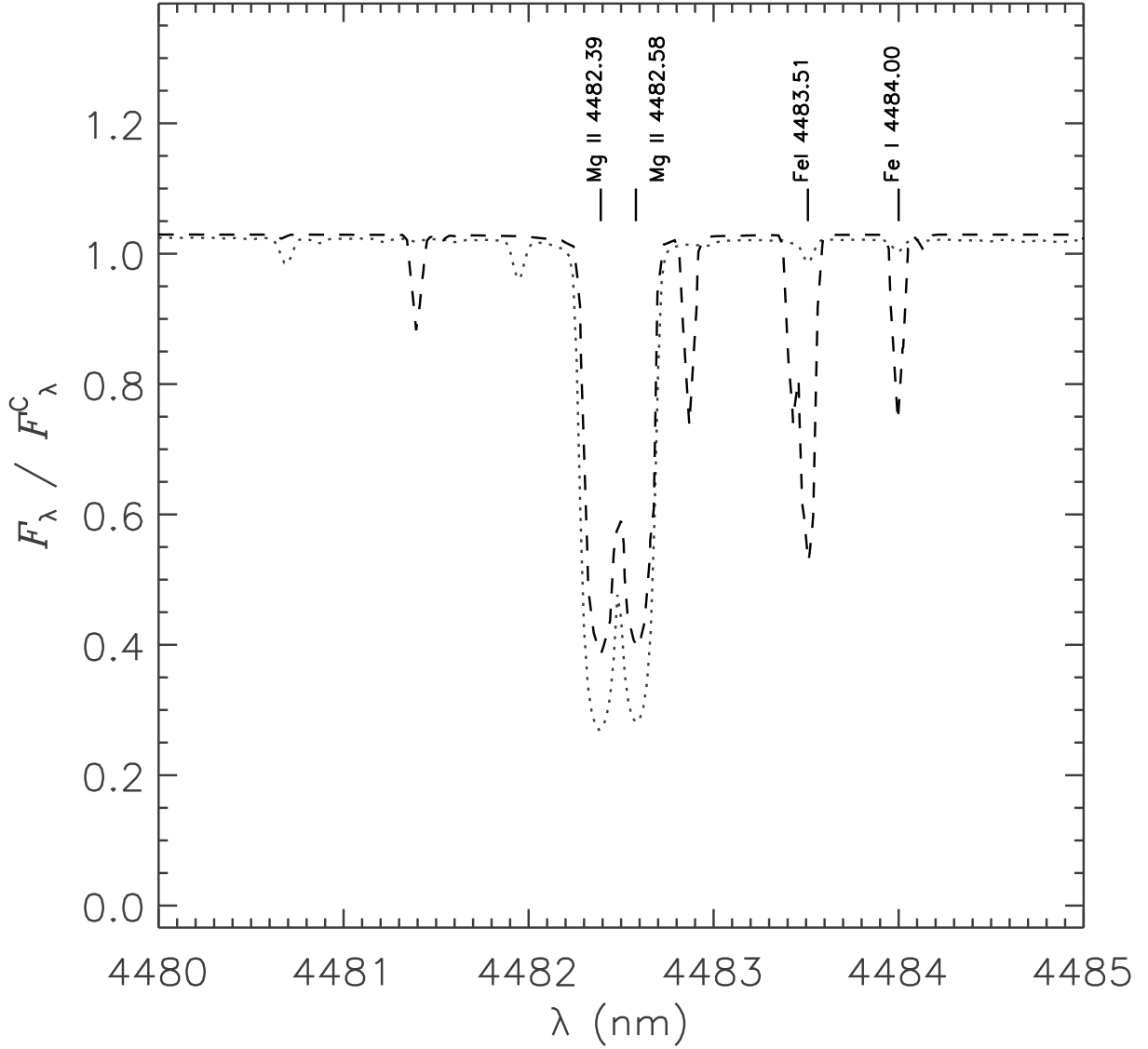}
\caption{Continuum rectified synthetic flux spectra, $F_\lambda(\lambda) / F_\lambda^{\rm c}(\lambda)$, in 
the region of the early-type MK classification diagnostic line \ion{Mg}{2} $\lambda$ 4481  
for a model of $T_{\rm eff}=10000$ K, $[{{\rm Fe}\over {\rm H}}]=0.0$, $\log g=4.0$,
and $\xi_{\rm T}=2.0$ km s$^-1$ as computed with GSS (dashed line) and 
Phoenix (dotted line).  
  \label{gssFlux10000}
}
\end{figure}

 Figs. \ref{gssFlux500045} and \ref{gssFlux500025} show a comparison of spectra in the 
important \ion{Ca}{2} $HK$ region as computed by GSS and Phoenix for models of
($T_{\rm eff}$/$\log g$/$[{{\rm Fe}\over {\rm H}}]$/$\xi_{\rm T}$) equal to (5000 K/4.5/0.0/1.0 km s$^{\rm -1}$)
and (5000 K/2.5/0.0/1.0 km s$^{\rm -1}$).  Fig. \ref{gssFlux10000} shows the same comparison
in the region of the strongest early-type MK classification diagnostic other than the \ion{H}{1} Balmer
lines, namely the \ion{Mg}{2} $\lambda 4481$ line, for 
a model of (10000 K/4.0/0.0/1.0 km s$^{\rm -1}$) (a comparison of \ion{H}{1} Balmer lines
is not helpful because GSS does not treat the difficult linear Stark broadening that is necessary
to even approximately model the saturated \ion{H}{1} lines).  For the early type star it is
clear that we systematically overestimate the strength of weak spectral lines as a result of 
underestimating the value of $\kappa_\lambda^{\rm c}$ by neglecting metal $b-f$ extinction.  However,
the strong and saturated lines that serve as pedagogically important classification diagnostics are
reproduced well enough to be credibly demonstrative.  

\begin{figure}
\plotone{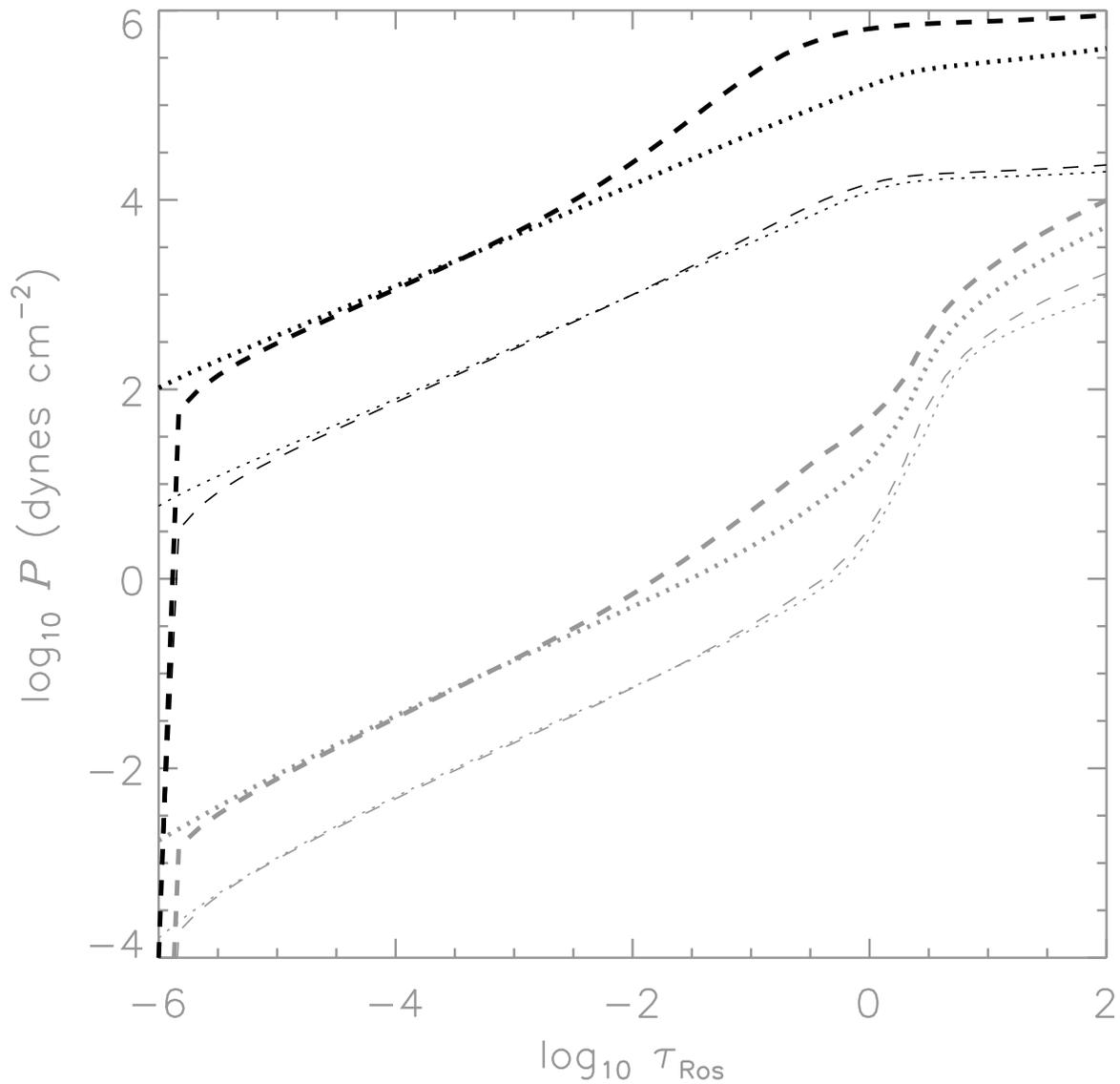}
\caption{The gas pressure, $P_{\rm gas}(\tau)$ (darker lines), and partial electron pressure, 
$P_{\rm e}(\tau)$ (lighter lines),
distributions for models of $T_{\rm eff}=5000$ K, $[{{\rm Fe}\over {\rm H}}]=0.0$, and $\log g$ equal to
4.5 (thicker lines) and 2.5 (thinner lines) as computed with GSS (dashed line) and 
Phoenix (dotted line). 
  \label{gssPe5000}
}
\end{figure}   

\begin{figure}
\plotone{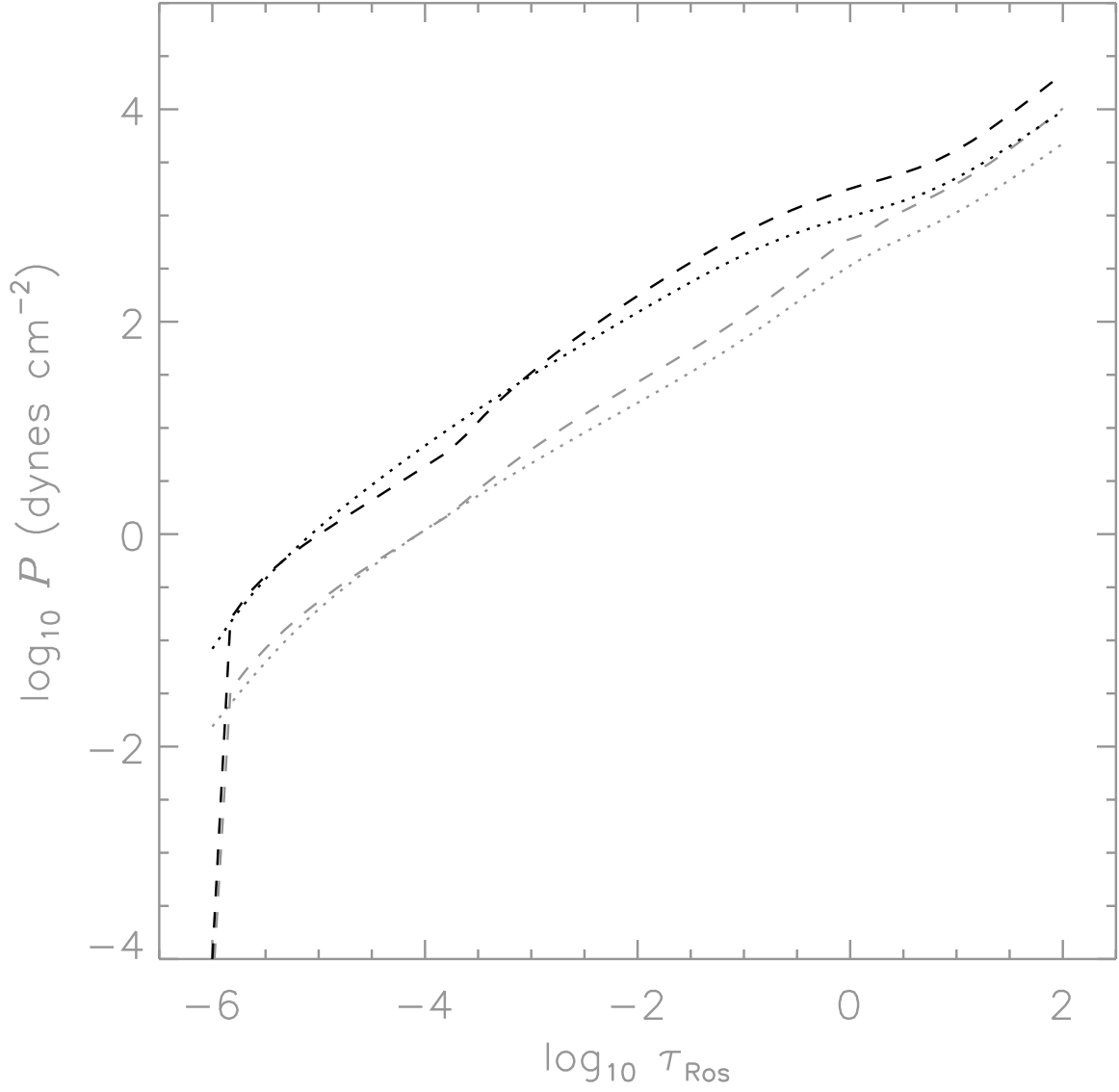}
\caption{Same as Fig. \ref{gssPe5000} but for a model of $T_{\rm eff}=10000$ K, $[{{\rm Fe}\over {\rm H}}]=0.0$, 
and $\log g = 4.0$.
  \label{gssPe10000}
}
\end{figure}

\begin{figure}
\plotone{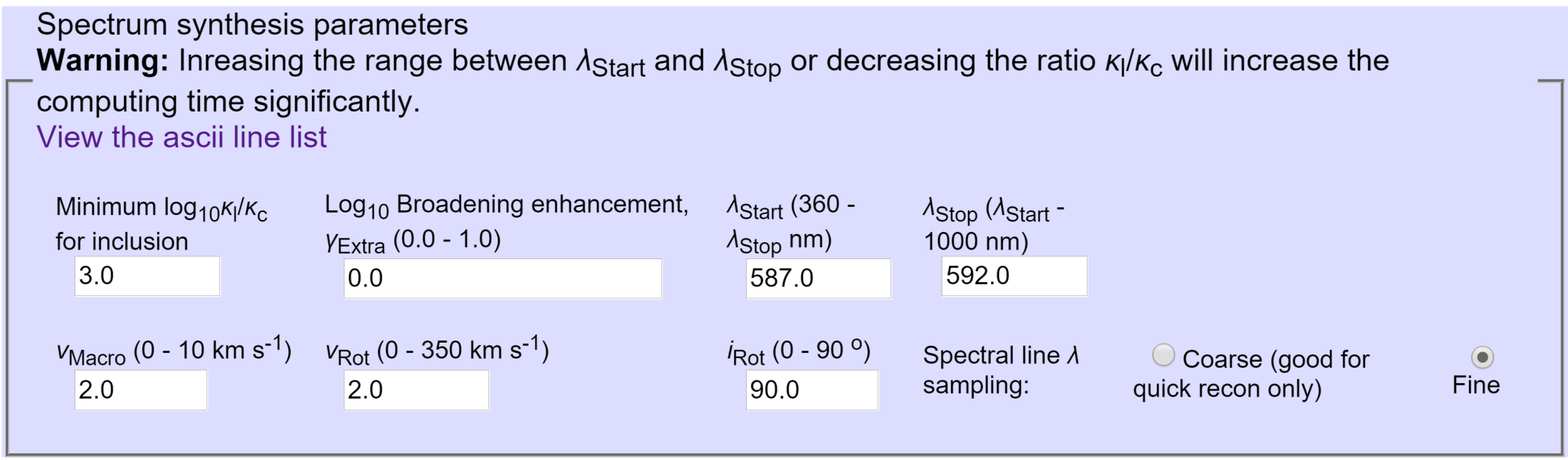}
\caption{The GSS UI input panel that controls the spectrum synthesis.  The
user may select the minimum ratio of
$\kappa^{\rm l}/\kappa^{\rm c}$ for inclusion in the synthesis, the value of 
the universal Lorentzian broadening enhancement, $\gamma_{\rm extra}$, the beginning and ending 
$\lambda$ values ($\lambda_{\rm start}$ and $\lambda_{\rm stop}$), whether the
$\lambda$ sampling of each line is ''fine'' or ''coarse'' (the latter is recommended
for quick reconnaissance work only), the RMS value of the macroturbulent velocity distribution,
$v_{\rm Macro}$, the surface equatorial rotational velocity, $v_{\rm Rot}$, and the inclination
of the rotational axis, $i_{\rm Rot}$.  Note the rest of the UI is similar to that of 
GS3, and is not shown in this report. 
  \label{finput}
}
\end{figure}

\begin{figure}
\plotone{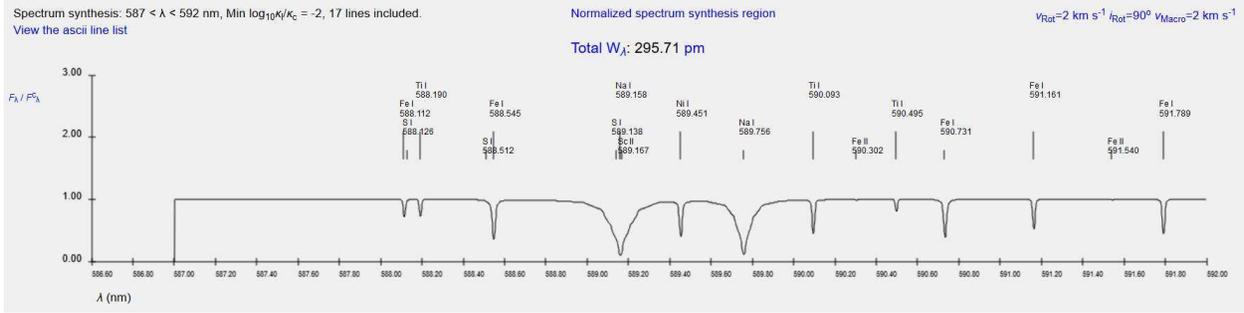}
\caption{A portion of the GSS UI output panel displaying the synthetic spectrum in the
vicinity of the \ion{Na}{1} $D$ lines for the Sun
with line identification labels.  The number of lines included in the synthesis region,
the total $W_\lambda$ value of {\it all} 
absorption in the synthesis region in pm, and the values of $v_{\rm Macro}$, $v_{\rm Rot}$
and $i_{\rm Rot}$ are all displayed.  The vertical line labeled 
''Filter'' indicates
the $\lambda$ value of the user-tunable narrow-band Gaussian disk image, $\lambda_{\rm disk}$
 (the same as
that of the GS3 UI and not shown here) - the $\lambda_{\rm disk}$ indicator is displayed 
if $\lambda_{\rm disk}$ falls within the synthesis region 
($\lambda_{\rm start} < \lambda_{\rm disk} < \lambda_{\rm stop}$) so that users can relate  
the appearance of the monochromatic disk to the amount of total extinction, 
$\kappa^{\rm l}_\lambda + \kappa^{\rm c}_\lambda$,
at $\lambda=\lambda_{\rm disk}$.
Note the rest of the UI is similar to that of
GS3, and is not shown in this report.
  \label{foutput}
}
\end{figure}

\section{Client-side post-processing and UI in JavaScript and HTML \label{Client}}

\paragraph{}

  GSS uses a web browser-based UI that is adapted from that of GS3, and \citet{graystar3} contains
a description and justification of the UI content, organization, and functionality.  The client
may be found at \url{www.ap.smu.ca/$\sim$ ishort/OpenStars/GrayStarServer}.  Post-processing
of the raw $I^{\rm +}_\lambda(\tau=0, \theta)$ and monochromatic surface flux, $F_\lambda(\tau=0)$,
 distributions to compute synthetic observables such as photometric color indices,
the $BVR$ white light and narrow-band Gaussian filter (see below) disk images, and the
equivalent width, $W_\lambda$, of absorption features is still
 performed on the client side in JS to keep the structure of the data sent from the 
server to the client as simple as possible, and to retain flexibility in how different
clients may choose to post-process basic modeling output from the server.  The JS client
also performs the macro-turbulent and rotational broadening of the synthetic spectrum
by convolution of the disk integrated $F_\lambda$ spectrum with appropriate broadening
kernels.  GSS improves upon the GS3 UI by making
use of the HTML5 $<$canvas$>$ element.  This allows for proper line plots, and 
color and brightness gradient shading in the rendering of the stellar disk and visible
flux spectrum, leading to greater photo-realism. 

\paragraph{}

The GS3
UI panel that contains the inputs for the atomic and line transition parameters of a generic,
representative, high resolution line profile are obviated here, and is replaced in GSS with 
a panel with inputs controlling the spectrum synthesis, shown in Fig. \ref{finput}.  These are 
the minimum
value of $\kappa^{\rm l}(\lambda_0)/\kappa^{\rm c}(\lambda_0)$ for a line to be included, 
the beginning and
ending wavelengths [$\lambda_{\rm start}$, $\lambda_{\rm stop}$], the RMS value of the 
macroturbulent velocity distribution ($v_{\rm Macro}$), the surface equatorial rotational
velocity ($v_{\rm Rot}$), and the inclination of the rotation axis with respect to the line-of-sight
($i_{\rm Rot}$),  
and a radio switch that selects for ''Fine'' or ''Coarse'' $\Delta\lambda$ sampling of the
line profiles.  To avoid synthesis calculations that would exceed the normal time limit of
an HTTP request (see below), especially with synthesis in the crowded blue spectral region, 
the value of $\lambda_{\rm stop}$ is limited to being less than
10 nm greater than that of $\lambda_{\rm start}$.  The ''Fine'' setting specifies nine points in 
the Gaussian core (a line center 
point and four points per
half-core distributed symmetrically about $\lambda_0$), and 18 in the Lorentzian wings
(nine per wing), for a total of 27 $\Delta\lambda$ points per line.  For the ''Coarse''
setting, these numbers are five and six, respectively, for a total of 11
$\Delta\lambda$ points per line.  Given the sparse $\lambda$ sampling in which
generally only spectral lines are sampled, the ''Fine'' setting is generally
necessary for a spectrum that would be visually meaningful to the inexperienced.  
The ''Coarse'' setting speeds up the calculation somewhat, and is more useful for quick 
reconnaissance.  
The UI also has a link that allows the user to view the plain text ASCII
version of the line list. 
Like GS3, GSS is equipped with presets for sample spectral 
lines that are pedagogically important.  In the case of GSS, choosing a preset causes
the range $[\lambda_{\rm start}, \lambda_{\rm stop}]$ to center on the preset line.  

\paragraph{}

  The simple monochromatic imaging filter that GS3 has has been replaced by a proper narrow band
Gaussian filter for which the user can adjust the value of the band-width, $\sigma$, as well
as tune the central wavelength, $\lambda_{\rm disk}$, as before.  The ability to adjust
the value of $\sigma$ leads to a wider variation of the limb darkening of the filtered 
disk image when $\lambda_{\rm disk}$ lies between $\lambda_{\rm start}$ and $\lambda_{\rm stop}$
in a region where many spectral lines are closely spaced.  Additionally, beyond what GS3
originally provided for, GSS includes the option to print out the individual 
chemical abundances on the logarithmic $A_{\rm 12}$ scale (the abundance distribution is
the solar one of \citet{grevesse10}), and to display and print out the total populations,
$\log N_{\rm k}$,
at the depth of $\tau_{\rm Ros}=1$ for the first
three ionization stages, $k$, of any species included in the synthesis. 
 
\paragraph{}


  GSS features an output panel, shown in Fig. \ref{foutput}, containing the synthetic spectrum, that is triple the width
of the standard output panels to enhance visualization of fine detail.  The data 
structure returned by the server includes the species identifications and 
$\lambda_0$ values for the lines included in the synthesis, and these are marked
on the plot.  GSS computes and displays the total equivalent width, $W_\lambda$, in pm, of {\it all}
lines included in the synthesis region.  To compute $W_\lambda$ for a single line
of interest, the user should restrict the range [$\lambda_{\rm start}$, $\lambda_{\rm stop}$]
to bracket and isolate the line.  This panel also displays the values of the macroscopic
broadening parameters $v_{\rm Macro}$,
$v_{\rm Rot}$, and $i_{\rm Rot}$, as specified by the user.  
 The GSS UI inherits from GS3 a monochromatic rendering of the
spatially resolved disk for which the user can tune the value of the imaged wavelength,
$\lambda_{\rm disk}$ (see \citet{graystar3}), and now, also, the RMS band-width, $\sigma$.  
If the value of $\lambda_{\rm disk}$ falls 
within in the spectrum
synthesis range ($\lambda_{\rm start} < \lambda_{\rm disk} < \lambda_{\rm stop}$), 
it is indicated in the spectrum synthesis plot.  This enables a user
to relate the appearance of the monochromatic disk to the amount of total extinction, 
$\kappa^{\rm l}_\lambda+\kappa^{\rm c}_\lambda$, near $\lambda = \lambda_{\rm disk}$. 
GSS inherits from GS3 a direct image of the visual band spectrum, which, in the case of
GSS only reflects those spectral lines that are contained in the range 
[$\lambda_{\rm start}, \lambda_{\rm stop}$].

\subsection{Rotational broadening and limb darkening coefficients (LDCs)}

 To compute a rotational broadening kernel as described by \citet{dfg3}, GSS
requires a narrow band or monochromatic continuum limb darkening coefficient (LDC), $\epsilon_\lambda$,
for the spectral region being synthesized, where for the linear limb darkening law
   
\begin{equation}
I^{\rm c}_\lambda(\theta)/I^{\rm c}_\lambda(0) \approx 1 - \epsilon_\lambda + \epsilon_\lambda\cos\theta .  
\end{equation}

 For a $\lambda$ value approximately midway
between $\lambda_{\rm start}$ and $\lambda_{\rm stop}$, GSS computes the mean $\epsilon_\lambda$ 
value for each of the 11 $\theta$ points from

\begin{equation}
\epsilon_\lambda = (I^{\rm c}_\lambda(\theta)/I^{\rm c}_\lambda(0) - 1) / (\cos\theta - 1) .   
\end{equation}

GSS includes the option to print
out the LDC values for all $\lambda$ values in the coarse $F^{\rm c}_\lambda$ grid.

\section{Client-server considerations \label{ClientServer}} 

\subsection{Client-side}

  The JS and HTML code running on the client sends the input modeling parameters to 
the server attached to an asynchronous HTTP request using the 
XMLHttpRequest() method.
The 'POST' and 'asynchronous=true' options are specified in the 
open() method of the object returned by XMLHttpRequest().
Calling the open() method in asynchronous
mode allows the JS and HTML code to proceed with any processing that it can while
waiting for the response from the server, which allows for a smoother user experience
given the time required to respond to more demanding spectrum synthesis parameters.
 
\paragraph{}

 
  The server returns the following data, in cgs units, 

\begin{enumerate}

\item{The vertical 
atmospheric structure, $T_{\rm Kin}(\tau)$, $P_{\rm Gas}(\tau)$, $P_{\rm Rad}(\tau)$,
 $\rho(\tau)$, $N_{\rm e}(\tau)$, $\mu(\tau)$, and $\kappa_{\rm Ros}(\tau)$. }

\item{The overall low resolution 
synthetic flux spectrum (SED) computed with and without line opacity, 
$\log F_\lambda(\log\lambda)$ and $\log F^{\rm c}_\lambda(\log\lambda)$.}

\item{The specific intensity distribution,
$\log I_\lambda(\log\lambda, \theta)$.}

\item{The high resolution continuum rectified spectrum in the
synthesis region, $\log F_\lambda(\log\lambda) - \log F^{\rm c}_\lambda(\log\lambda)$.}

\item{The chemical species identifications and central wavelength, $\lambda_0$, of
all lines for which extinction was included in the synthesis.}

\item{The number of spectral lines for which extinction was included in the synthesis.}

\item{The number of
vertical $\log\tau$ points sampling the atmosphere, $\log\lambda$ points sampling the SED,
$\theta$ points sampling the $I_\lambda(\theta)$ distribution, and $\lambda$ points sampling the
synthesis region.}

\item{The monochromatic linear LDC values, $\epsilon_\lambda$.}

 and 
\item{The total ionization stage population values, $\log N_{\rm k}$.}

\end{enumerate}

All numeric quantities are transmitted as their natural (base $e$) logarithms.  This 
data arrives encoded in a lengthy JSON (JavaScript Object Notation) string and is 
accessed with the responseText field of the object returned by XMLHttpRequest(). 
The JS code decodes the JSON string with the JSON.parse() method, and the data accessed
through the corresponding fields of the object returned by JSON.parse().  From this point
the JS code proceeds to display the modeling results in the same way GS3 does, with the
addition of the spectrum synthesis content.

\subsection{Server-side}

  The HTTP request sent by the client is received by a PHP script
and the attached model parameters are accessed through the corresponding keyed elements
of the intrinsic superglobal $\mathdollar\mathunderscore$POST[~ ] array variable.  The parameters are ''sanitized'' to protect the
server from malicious cross-site scripting (XSS) and appended as command-line arguments to an invocation,
made with the exec() function, of the executable code that performs the modeling.
The PHP script captures everything written to the standard output (stdout) by the executable, concatenates
it in one lengthy string variable, and encodes it in JSON for return to the client. 

\subsubsection{Open standards for spectrum synthesis in the web era \label{standards}}

  The structure of the JSON string used by GSS to transmit the modeling output
from the server to the client may be viewed by visiting the URL
\url{www.ap.smu.ca/$\sim$ ishort/OpenStars/GrayStarServer/solarDemo.php}. 
This string represents a suggested standard for marking up and packaging 
spectrum synthesis and model atmosphere data for both transmission and
storage.  The advantage of using JSON is that any application written in a 
web-aware programing language can simply call a library routine ({\it eg.} 
the JSON.parse() method of JS) to decode the JSON 
string, unpack the variables, and detect the name for each field
containing a data item.    

\paragraph{}

  This suggested standard implies a number of more fundamental suggestions for 
standard practice that involve expanding the definition of a ''synthetic 
spectrum'' to necessarily and universally include the natural logarithm of the
following data items in pure cgs units:


\begin{enumerate}

\item{ The $\lambda$ and 
corresponding $F_\lambda(\tau=0)$ values of the blanketed spectrum, as is currently the expectation, 
{\it along with} those for the unblanketed continuum spectrum, $F^{\rm c}_\lambda(\tau=0)$, computed 
with the same atmospheric model,}

\item{ The corresponding $I^{\rm +}_\lambda(\tau=0, \theta)$ values at a set of $\theta$ values that 
critically sample the $\theta$ dependence of $I^{\rm +}_\lambda$ (typically 16 values),}

\item{ The chemical identities and $\lambda_0$ values of all lines for which extinction
was included in the spectrum synthesis calculation,}

\item{ The vertical structure of
the atmospheric model that produced the spectrum, including $\tau_{\rm Rosseland}$,
one of $\tau_{200}$, $\tau_{500}$, or $\tau_{1200}$ as appropriate, and the corresponding 
$T_{\rm kin}$, $P_{\rm Gas}$, $\rho$, $\Delta r$, $N_{\rm e}$, and mean molecular weight
($\mu$) structures,}

\item{ The stellar parameters $T_{\rm eff}$, $\log g$, $[{A\over H}]$ (if scaled solar
modeling is used), $\alpha$-element enhancement, $\xi_{\rm Microturbulence}$,
and one of $M$, $L_{\rm bol}$, or $R$ if the modeling uses spherical geometry,} 


\item{The values of the individual abundances of the astrophysically important 
chemical species,}

and that 

\item{The community use standard labels for each of these data items in the JSON mark-up.}

\end{enumerate}

\paragraph{}

Item 1) will ensure that end-users have a way to accurately and unambiguously rectify 
synthetic spectra.  Item 2) will become increasingly important as IR- and visible-band stellar interferometry
become increasingly important and the community wishes to model visibility and imaging
data.  Item 3) would be especially valuable for identifying any ambiguities or disagreements
about the identity of the species most responsible for any given spectral feature.  Items 4)
and 5) would make explicit the atmospheric structure responsible for a given spectrum and serve
as internal consistency checks, and we note
that the $\tau_{\rm Rosseland}$ and $\Delta r$ values can be combined with the 
$\rho(\tau_{\rm Rosseland})$ structure to approximately recover the mean extinction coefficient
structure, $\kappa_{\rm Rosseland}(\tau_{\rm Rosseland})$.  
Item 7) would require that 
the community adopt universal variable names for all these data items in any applications
that are developed to work with synthetic spectra, and would increase the transparency
necessary for independent critical scrutiny of codes.  This suggested standard 
involves incorporating atmospheric models into the definition of a ''synthetic spectrum''.
The current wide-spread availability of high band-width, and of large computer memory and data storage 
capacity, provides us with an opportunity adopt a standard that will maximize transparency
and minimize inadvertent misinterpretation of a synthetic spectrum. 
 
\subsection{Extensibility}

  The executable atmospheric modeling and spectrum synthesis code on the server side could
be any code compiled from any programing language.  As long as the I/O routines have been adapted so
that the code expects the command line arguments that the server-side PHP script includes in the
exec() call, and so that the code writes to stdout the data items in the order and format the PHP
script expects to capture, then GSS will 
work just as it does with its current Java modeling code.  Therefore, if responsiveness is not a 
concern, The GSS client 
could be used to interact with, and display the results of, research grade codes written in 
scientific programing languages such as FORTRAN and C.  As a result, results of research quality
could be displayed with the platform-independent client-side UI.

\section{EPO \label{EPO}}


Beyond the demonstrations enabled by GS3, the spectrum synthesis plot could be used in
conjunction with the monochromatic disk image to demonstrate or investigate how the
monochromatic image changes as one scans with $\lambda_{\rm disk}$ through the wide 
range of total monochromatic
extinction, $\kappa^{\rm l}_\lambda + \kappa^{\rm c}_\lambda$, in a narrow 
$\lambda$ range provided by a 
saturated spectral line.  Such
a demonstration is aided by the $\lambda_{\rm disk}$ indicator (vertical line
labeled ''Filter'' in Fig. \ref{foutput}) in the spectrum
synthesis plot.  A classic 
example is to scan through the blue wing of the heavily saturated \ion{Ca}{2} K line
in a late type dwarf or giant.
The GSS interface allows students to directly see how the monochromatic limb-darkening
decreases as one progressively scans from the far wing to line center at $\lambda_0$.
This serves as an important demonstration of the Eddington-Barbier relation in
$\lambda-$ and $\theta$-space simultaneously, and is
relevant to how the vertical structure of the
solar atmosphere can be determined semi-empirically.  An animation of the this
demonstration in the vicinity of the \ion{Mg}{1} $b$ line is available at
www.ap.smu.ca/$\sim$ishort/OpenStars/MgIbScan.gif.

\paragraph{}

 GSS can be used to directly demonstrate the dramatic increase in the density of spectral 
lines per $\Delta\lambda$ interval (an important phenomenon to consider when planning
spectroscopic observations), and students could be asked to draw a mathematically based 
conclusion about the variation in the density of lines per energy interval, $\Delta E$.
The synthetic spectrum is annotated with line identifications, and students can 
be asked to think about the relative representation of different
elements among the spectral lines, appreciate by direct experience the disproportionate
representation of Fe, and be asked to think about what peculiarities of Fe give
rise to the phenomenon (both relatively large $[{{\rm Fe}\over {\rm H}}]$ and a rich
energy level structure).  Students could be asked to note which
ionization stages are represented among the spectral lines in stars of various
$T_{\rm eff}$ over a wide range, and to relate what they find to ionization
equilibrium and MK spectral classification.  The user may inspect the plain text
ASCII version of the line list by clicking on the link provided in either the
input or output panels for the spectrum synthesis, and relate the atomic data values
such as those of $f_{\rm ij}$ and $\chi_{\rm i}$ to the strength of features in
the synthetic spectrum. 

\section{Research \label{Research}}

  Like GS3, GSS makes use of a method of estimating the background continuous extinction,
$\kappa^{\rm c}_\lambda$,
that is much more approximate than that of research grade codes (an {\it ad hoc} 
combination of Kramer's scaling laws for the different major extinction source types,
see \citet{graystar3}).
Computed line strength is sensitive to $\kappa^{\rm l}_\lambda/\kappa^{\rm c}_\lambda$,
so the modeled line strengths are not of research accuracy.  However,
the line strengths {\it are} sufficiently accurate that GSS could serve as a reconnaissance 
tool to remind oneself quickly whether a particular chemical species of interest has
clean lines of useful strength  in a particular $\lambda$ range, and this could be useful in guiding decisions
that need to be made quickly ({\it eg.} if one has unexpected time to re-tune a 
spectrograph during an observing run), or when one is writing an observing proposal.   

\paragraph{}

  As noted above, the GSS client-server interaction and client-side UI are
oblivious to the nature of the executable that performs the server-side
physical modeling, and it could be a research-grade code, particularly
if the server is powerful enough to execute a research-grade code quickly.
This would allow the GSS client UI to be used for inspection of research
results.  The ease of producing client-side applications to post-process
and display research-grade modeling results would be enhanced by conformity
to the standards for structuring and marking up spectrum synthesis products
for transmission and storage that were described in Section \ref{standards}.      

\section{Conclusions and future work \label{Conclude}}

  GSS extends the functionality of GS3 by providing for spectrum synthesis so that
plausible spectra can be displayed for EPO purposes.  To achieve this, it moves
beyond the pure client-side modeling of GS3 by moving the physical modeling
to the server side and developing a standard for the transmission of 
atmospheric modeling and spectrum synthesis data via the web using the XMLHttpRequest()
method.  This moves the web-based astrophysical modeling approach pioneered 
by GS3 into a realm that allows for more powerful and realistic modeling 
with any language that be compiled on the server side.  

\subsection{Future work}

  We encourage others who are interested to become involved in the following:

\paragraph{Modeling}

  A natural next step would be to include molecular line opacity in the spectrum
so that GSS is even more useful for pedagogically demonstrating and investigating late-type 
stars (and possibly brown dwarfs and ''hot Jupiters'').  Doing so will be a challenge 
given the constraints of the low-performance 
computational environment that GSS is designed for, but the ''just overlapping line
approximation'' (JOLA) may be efficient enough to be useful.  Relatedly, and in 
the spirit of the GrayStar project, we plan to investigate the possibility of
implementing the line list as a Structured Query Language (SQL) database rather than 
as a byte file, which may allow for faster read times and enable inclusion
of molecular lines.  An SQL-based spectrum synthesis could also allow for more direct 
inspection and perturbation of atomic data while preforming the synthesis calculations.
 The next step in improving the treatment of the background continuous opacity, $\kappa^{\rm c}_\lambda$,
in the blue and near-UV bands would be to add extinction caused by $b-f$ transitions of 
key metals.  These extinction sources are generally non-hydrogenic, and a way to treat them
within the special performance constraints of GSS would be required.
The improvements to the atmospheric modeling detailed in Section \ref{newphysics} can now be
back-propagated to the GS3 JS code to improve its realism.  
Generally, both the server-side modeling and
the client-side post-processing and UI rendering might benefit from hardware acceleration
using the the server's and client's graphics processing units (GPUs).  Both Java
and the HTML $<$canvas$>$ element provide for access to the OpenGL library (JOGL
and WebGL, respectively), and the faster processing time achievable might allow 
for more realistic responsive modeling.

\paragraph{EPO}

  A useful supplement to GSS (and GS3) would be a set of tested procedures and discussion questions
for lab-style homework assignments that instructors could adopt and adapt, and we plan to develop these
and make them available as we gain experience.  It would be interesting to have both
qualitative and quantitative
assessments of the pedagogical efficacy of GSS (and GS3) based on student performance and 
surveys, and we plan to pursue this as well and make the results available to the
community.



\acknowledgments
The author acknowledges Natural Sciences and Engineering Research Council of 
Canada (NSERC) grant RGPIN-2014-03979, and David F. Gray for helpful
private communications about computing extinction sources.

\clearpage



\clearpage






\end{document}